# Accurate phase retrieval of complex point spread functions with deep residual neural networks

LEONHARD MÖCKL,[1] PETAR N. PETROV,[1] AND W.E.MOERNER[1,*]

[1]*Department of Chemistry, Stanford University, Stanford, CA 94305, USA*
*\*wmoerner@stanford.edu*

**Abstract:** Phase retrieval, i.e. the reconstruction of phase information from intensity information, is a central problem in many optical systems. Here, we demonstrate that a deep residual neural net is able to quickly and accurately perform this task for arbitrary point spread functions (PSFs) formed by Zernike-type phase modulations. Five slices of the 3D PSF at different focal positions within a two micron range around the focus are sufficient to retrieve the first six orders of Zernike coefficients.

## 1. Introduction

Wavefronts carry two fundamental types of information: (i) Intensity information, i.e. photon flux, and (ii) phase information. Detectors typically used in optical measurements rely on the conversion of the incoming fields to electrons. Due to the physics of the detection process, without interferometry only the intensity information of the incoming wavefront can be recorded whereas the phase information is lost. However, it is possible to extract useful phase details to some extent from the recorded intensity information. This task has been coined the phase problem, an issue of fundamental importance not only in optics, but in many other areas of physics, e.g. X-Ray crystallography, transmission electron microscopy, and astronomy [1-3].

The process of solving or approximating the phase problem is generally termed phase retrieval (PR), and numerous PR algorithms have been developed. Typically, they involve iterative optimization for the phase information under the constraints of the known source and target intensities as well as the propagating function [4-9]. For example, a set of images of a known point source at various axial positions has been used to estimate pupil phase. A recent application of this approach has used PR for the design of tailored phase masks for 3D super-resolution imaging [10]. While these approaches allow accurate estimation of the phase information, they are computationally demanding and, as a result, relatively slow. Moreover, they require oversampling of the feature space, i.e. a small increase in extracted phase information requires a large amount of additional intensity information [11, 12].

Deep neural nets (NNs) have recently been demonstrated to be useful tools in optics and specifically in single-molecule imaging [13, 14]. In these approaches, a learning process based on known image inputs trains the coefficients and weights of the NN, and the NN then processes unlearned images to extract position information or other variables. Especially relevant for PR, it has been shown that a residual neural net (ResNet) [15] is capable of extracting wavefront distortions from biplane PSFs, which could be efficiently used to correct for aberrations with adaptive optics [16]. Also, it was demonstrated that deep learning can be used to recover images at low-light conditions [17] and to accelerate wavefront sensing [18-20]. These findings led us to consider whether an approach based on a carefully chosen NN architecture might be able to tackle the fundamental problem of PR of arbitrary PSFs. In our design, after learning, the NN performs PR on a small set of PSFs, requiring nothing as input but a PSF stack from a range of axial positions and directly returning the phase information as Zernike coefficients, as schematically depicted in Fig. 1(a).

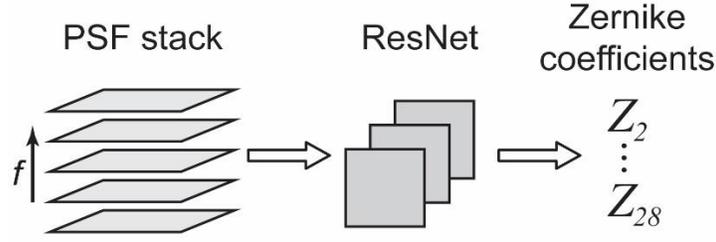

**Fig. 1.** Workflow. A stack with a few slices of the PSF at different focal positions *f* is supplied to a deep residual neural net which processes the images and directly returns the Zernike coefficients of order 1 to 6 (Noll indices 2 to 28) that correspond to the phase information encoded in the PSF images. The first Zernike coefficient, piston, does not transport phase information and is therefore not included.

## 2. PSF simulations

A cornerstone of any approach involving NNs is a sufficiently large dataset for training of the NN as well as an independent validation dataset to verify that the trained NN is unbiased. To provide our NN with training and validation data, we turned to accurate PSF simulations. PSFs were simulated by means of vectorial diffraction theory [21, 22]. For training, we simulated the PSFs at focal positions -1, -0.5, 0, 0.5, and 1 µm. The simulated emitter was positioned directly at a glass coverslip in index-matched media. Phase information was introduced by multiplying the Fourier plane fields by a Zernike phase factor with random Zernike coefficients of order 1 to 6 (Noll indices 2 to 28) to values between $-\lambda$ and $\lambda$. Typically, the higher-order Zernike coefficients tend towards zero in experimental settings, but as it was our goal to sample the parameter space equally, we did not impose any such limitation on the sampling within this set. The zero-order Zernike coefficient does not transport phase information and was hence not considered. Furthermore, we included camera specifications typically encountered with EMCCD detectors (with the exception of excess noise), signal and background photons, and Poisson noise. The relevant parameters are summarized in Table 1. Naturally, the chosen parameters are somewhat specific for a particular situation and can be changed according to the requirements of a specific problem.

**Table 1. PSF simulation parameters.**

| Parameter | Value |
| --- | --- |
| ROI size | 25x25 px |
| pixel size | 117.2 nm |
| signal photons | 12,500 – 25,000/PSF |
| background photons | 50 – 150/px |
| mean no-light counts | 101 |
| σ(no-light counts) | 4.15 |
| No-light counts noise model | Gaussian |
| photon noise model | Poisson |
| conversion gain | 26.93 count/photoelectron |
| EM gain | 186 |
| z_emitter | 0 nm |
| λ_emission | 671 nm |
| dipole contributions in x, y, z | (1,1,1)/3 |
| NA_objective | 1.4 |
| n_coverslip | 1.518 |
| n_medium | 1.518 |
| # of Zernike coefficients | 27 (order 1 to 6) |
| Zernike coefficient range | $-\lambda$ to $+\lambda$; uniformly sampled |

Fig. 2(a) depicts three representative PSFs at the five simulated focal positions. The resulting PSFs display complex shapes and rapid changes in their appearance when the focal position is altered; an expected behavior given the variations as large as $2\lambda$ in the Zernike coefficients up to high-order as visible in Fig. 2(b).

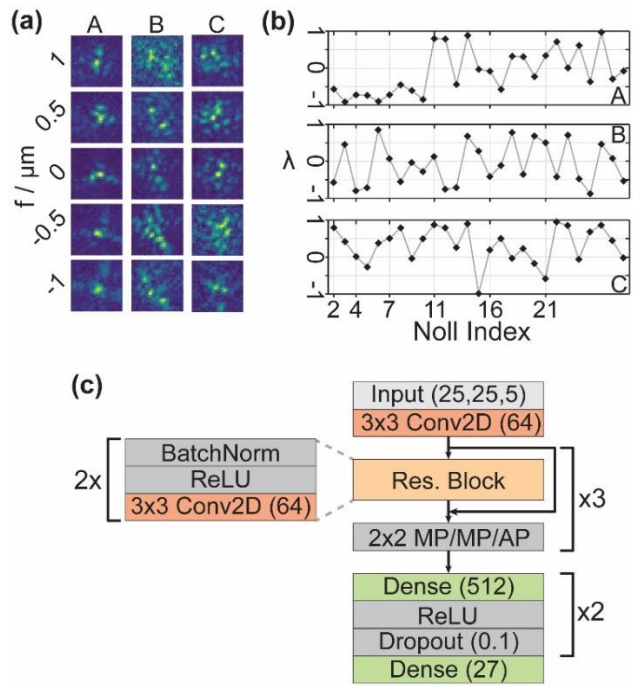

**Fig. 2.** (a) Three representative PSFs (A, B, C) at focal positions from -1 to 1 µm. (b) Zernike coefficients for the three representative PSFs shown in (a). For clarity, only Noll indices that correspond to a change in the order $Z_n$ of the Zernike coefficients are marked along the x-axis. (c) Schematic NN architecture. The PSF stack is supplied to the NN as a 25x25 pixel image with five channels, corresponding to the five focal positions. After an initial 3x3 2D convolution layer with 64 filters, three residual blocks follow, each consisting of two stacks of batch normalizations, ReLU activation, and a 3x3 2D convolution layer with 64 filters. After each residual block, the output of the residual block and its respective input are added. Then, 2x2 pooling is performed (MaxPooling, MP, after the first two residual blocks and average pooling, AP, after the third). Finally, two fully connected layers with 512 filters follow, each with ReLU activation and a dropout layer (dropout rate = 0.1). The last layer returns the 27 Zernike coefficients of order 1 to 6.

## 3. Network architecture

With realistic PSF simulations readily available, we now describe the NN architecture and training. Two requirements identified during optimization of the parameters are worth mentioning: (i) Training set size. The ability of the net to extract the Zernike coefficients sharply decreases when the training set is too small. This was not a gradual process. For example, at a training set size of 200,000 PSFs, training was not successful. We used 2,000,000 training PSFs. (ii) Choice of architecture. Simple convolutional nets (ConvNets) in our hands were not able to return the Zernike coefficients. Presumably, a ConvNet with just few layers does not provide enough parameters to learn the complex phase information. However, a ConvNet that is deep enough to exhibit a sufficiently large parameter space likely suffers from the so called degradation or vanishing gradient problem [15]. Architectures such as the Inception [23] and Xception [24] networks, on the other hand, excel at image classification tasks because they are able to recognize features at very different length scales. Though these architectures have been described for wavefront sensing [19, 20], we suspected that these architectures could be improved for PR. Specifically, a single feature in a PSF image is by itself not necessarily indicative of the phase. Rather, the whole PSF image encodes the phase information. Indeed, the networks described in [19] and [20] are quite large, potentially due to a suboptimal architecture that requires increased size. In contrast, the simpler ResNet architecture we chose here seems to be well situated for the problem of PR. The key feature of ResNets are residual blocks which feature skip connections between earlier and deeper layers, allowing for residual mapping. Indeed, the architecture we chose is compact and, as a result, fast (analysis of the validation set with 100,000 PSFs took about eight minutes on a standard desktop PC without a high-end processor or GPU).

The network architecture is schematically depicted in Fig. 2(c). The PSFs are supplied to the NN as five-channel images, corresponding to the five focal positions. After an initial 2D convolution, three residual blocks with batch normalization, ReLU activation, and 2D convolution follow with two maxpooling and one averagepooling step afterwards, respectively. As the term "residual" indicates, the output $x$ of a layer $n$ (e.g. the output of the initial convolutional layer) is stored before being passed on to the next layer (e.g. the batch normalization layer of the first residual block). Later, the earlier output $x$ is added to the output $y$ of a deeper layer $m$ (e.g. the convolutional layer of the second residual block). The joint outputs $x+y$ are then passed as input to the next layer $m+1$. The output $x$ is therefore (i) passed to the next layer $n+1$ and (ii) also skips several layers. Finally, two fully connected layers are implemented before the final output layer, which returns the 27 Zernike coefficients of order 1 to 6. After each of the two fully connected layers, dropout at a rate of 0.1 is implemented to avoid overfitting [25].

The NN was implemented in Keras with Tensorflow backend and trained on a desktop PC equipped with 64 GB RAM, an Intel Xeon E5-1650 processor, and an Nvidia GeForce GTX Titan GPU. The relevant training parameters are summarized in Table 2. Convergence was reached after training for approximately 12 hours and 77 epochs.

Table 2. Training parameters.

| Parameter | Value |
| --- | --- |
| # of PSFs stacks for training | 2,000,000 |
| # of PSF stacks for validation | 100,000 |
| optimizer | Adam ($\beta_1$=0.9, $\beta_2$=0.999, $\varepsilon$=1E-8) |
| Loss function | MSE |
| initial learning rate (LR) | 0.002 |
| $\varepsilon$ for LR decrease | 0.0001 |
| factor for LR decrease | 0.5 |
| patience | 3 epochs |
| minimal LR | 1E-6 |
| batch size | 512 |

## 4. Results

We now assess the performance of the NN on blind validation images not present in the training set, created as above with random Zernike coefficients. Fig. 3(a) depicts the overall deviation in wavelength units between the predicted Zernike coefficients and their ground truth value for the validation dataset (all 27 values for the 100,000 PSFs are pooled). The deviations are symmetric and centered at zero, indicating that the NN does not exhibit a bias toward over- or underestimating. Also, the relatively small width of the histogram (standard deviation of approximately 0.24) indicates reasonably precise predictions.

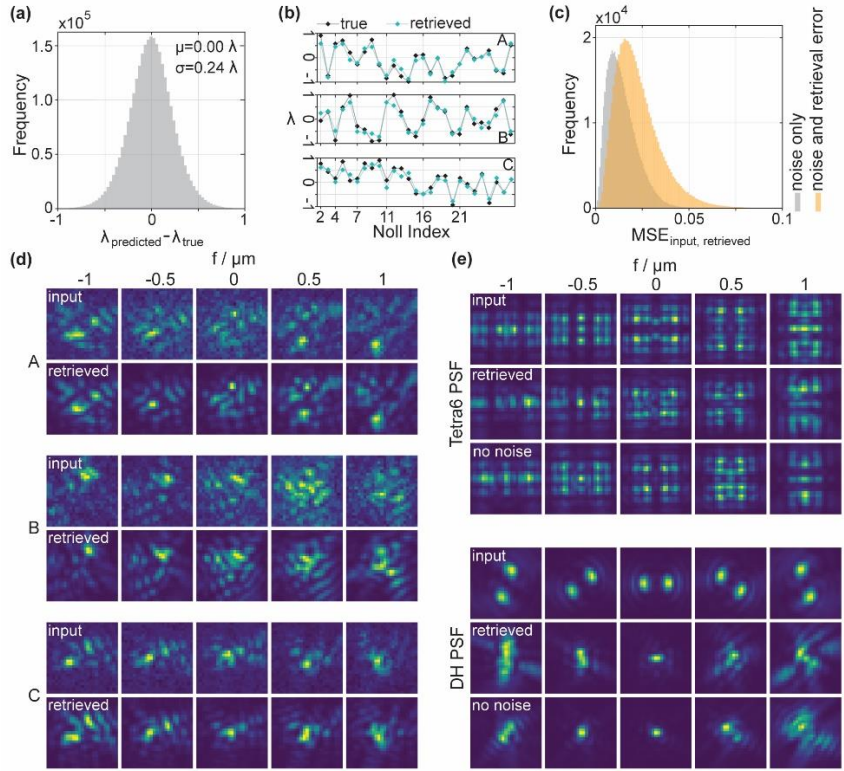

**Fig. 3.** (a) Histogram of the differences in wavelength units between the retrieved and the ground truth values for all Zernike coefficients in the validation dataset (100,000 PSF stacks with 27 coefficients each). (b) Representative examples for the agreement between the retrieved and ground truth Zernike coefficients for three PSFs. Notably, the prediction is accurate over the whole range of Zernike coefficients, up to the highest order. (c) Histogram of mean squared errors between input and retrieved PSFs, calculated pixelwise for each PSF slice of the 100,000 validation PSFs (orange, 500,000 slices total). Input and retrieved PSFs are normalized. The high agreement between the retrieved and true Zernike coefficients translates to low MSEs. Note that a significant fraction of the error stems from Poisson noise in the input PSFs. This is evident from the MSE between input PSFs without noise and input PSFs that include noise (gray histogram). (d) Comparison between the input PSFs at the five simulated focal positions and the PSFs generated from the retrieved Zernike coefficients, corresponding to the data plotted in (b). The agreement is very good. Note that the retrieved PSFs do not include background or Poisson noise whereas the input PSFs to be retrieved do. (e) Phase retrieval of the Tetrapod PSF with 6 µm range (Tetra6, top) and the double-helix (DH) PSF (bottom). Retrieval results from an additional NN are included which was trained on PSFs not including noise (labeled "no noise"). Note in this case, the input PSFs do not include noise as would typically be the case for a specific phase mask design. The DH PSF cannot be well represented using the Zernike basis set; and as a consequence, PR is poor. In contrast, the Tetra6 PSF can be represented using the Zernike basis set, leading to retrieved PSFs that approach the ground truth. As expected, the NN that was trained on PSFs without noise performs slightly better.

In agreement with the overall small deviations between prediction and ground truth, the NN was able to predict the Zernike coefficients corresponding to single PSFs accurately. This is shown for three representative random phase cases in Fig. 3(b) (see Figure 4 for more details).

The predicted Zernike coefficients are close to the ground truth values, but they are not perfect. Thus, we wanted to investigate next if prediction and ground truth agree on the level of the PSFs as well. For this, we performed PR on PSFs using the NN and then calculated the shape of the PSFs with the predicted Zernike coefficients. The results are depicted in Fig. 3(c) and Fig. 3(d). Fig. 3(c) shows the histogram of mean squared errors (MSEs) between input and retrieved PSFs, calculated pixelwise for each slice of all PSFs (orange). As is clearly visible, the MSE is low, indicating good agreement between input and retrieved PSFs at the level of the individual pixels. Notably, a significant part of the error is to be attributed to the Poisson noise in the input PSFs (compare to gray histogram). Fig. 3(d) shows the retrieved PSFs corresponding to the plots in Fig. 3(b) (compare plots A, B, C with PSFs A, B, C). As is evident from the comparison, the retrieved PSFs agree very well with the input PSFs. As expected from the MSE analysis, there are minor differences, but not only the overall shape, but also the intricate details of the complex PSFs are recovered at a high level of detail without the need to perform additional refinement using conventional PR algorithms.

To explore the utility of our PR approach, we asked whether the NN is able to perform PR on PSFs already used in praxis. Two of the most relevant PSFs for extracting 3D position are the double-helix (DH) PSF [26] and the Tetrapod PSF with 6 μm range (Tetra6) [27, 28]. Here we generated input PSFs under no-noise conditions to focus on the performance of the PR alone. The results are depicted in Fig. 3(e). Evidently, PR on the DH PSF is poor. This is fully expected as the Zernike basis set that we used to characterize the phase information is not suitable for the DH PSF, which is based on Gauss-Laguerre modes with singularities. Nevertheless, the NN retrieves a PSF that is vaguely related to the DH PSF. In contrast, the Tetra6 PSF can be described by the Zernike basis set, thus it is well retrieved. In this context, it should be mentioned that some Zernike coefficients were predicted to values well outside the training range of -λ to λ to retrieve the PSF, underscoring the robustness of the NN. Also, we note that including noise in training is not ideal when the NN is used for phase mask design. In fact, it is more reasonable to train the NN in noise-free conditions to just concentrate the information on the effect of the phase mask. This was confirmed by retrieving the Tetra6 PSF with a NN that was trained on noise-free PSFs, which yielded an improved result (labeled "no noise" in Figure 3(e)). Nevertheless, as the figure shows, also the NN that was trained on PSFs with noise still performs well for the Tetrapod.

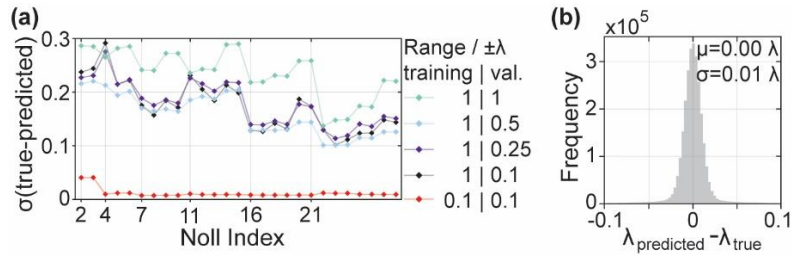

**Fig. 4.** (a) Standard deviations of the difference between predicted and ground truth Zernike coefficients for varying ranges. With decreasing range, the predictions of the Zernike coefficients first become more exact before plateauing. If a net is used that has been trained on a smaller range of Zernike coefficients ("trained", red curve), the predictions are very exact. This verifies that the plateauing solely originates from the wider range of Zernike coefficient values in the training set for the original NN. (b) Histogram of the differences between predicted and ground truth values for all Zernike coefficients for the NN trained with a coefficient range of ± 0.1 λ and a corresponding validation dataset (100,000 PSF stacks).

Finally, we investigated how the NN performs for varying ranges of Zernike coefficients. For this, we prepared additional validation sets with Zernike coefficients ranging between ± 0.5, ± 0.25, and ± 0.1 λ. The deviations between predicted and true Zernike coefficients decreased for a range of ± 0.5 λ compared to the full range ± 1 λ, but they did not further decrease for ± 0.25 and ± 0.1 λ (Fig. 4(a)). We suspected that this is due to the composition of the training set: When randomly picking 27 numbers between -1 and +1, the chances are very low that all of them fall into the range of ± 0.1 or even ± 0.25. In other words, PSFs corresponding to such small overall Zernike coefficients were underrepresented in the original training set. It is well known that NNs perform poorly when the parameter space covered by the training set is not optimal. To test this hypothesis, we trained a NN on PSFs corresponding to Zernike coefficients ranging between -0.1 and +0.1 λ ("trained", red curve in Fig. 4(a)). Indeed, this NN performed excellently on the validation data with the smaller range of ± 0.1 λ. This verifies that the plateauing observed for the original NN is primarily due to the wider range of Zernike coefficient values in the corresponding training set and can be easily overcome, dependent on the specific PR problem to be addressed.

It should be noted that the tip and tilt Zernike modes (Noll indices 2 and 3) do not change the shape of the PSF, but simply translate it along the vertical or horizontal dimensions. As such, in a real experiment they correspond to offsetting the PSF laterally with respect to the center of the chosen region of interest and cannot be measured directly. The high precision of the estimates of the remaining coefficients suggests that the NN is robust to such a lateral offset of the PSF and recovers the important aberration terms precisely and accurately (cf. Fig. 4(b) – note the very small histogram width).

## 5. Conclusion

In conclusion, we have developed a deep residual neural net that performs fast and accurate PR on complex PSFs. We investigated the net architecture and the training data parameters and verified the capability of our approach on realistic simulations of complex PSFs carrying Zernike-like phase information. We also demonstrated that the NN is able to perform accurate PR on the experimentally relevant Tetra6 PSF. From this, it will be straightforward to expand this approach to PR of non-Zernike-like phase information using different basis sets and to transfer the residual net concept to more applied tasks such as phase mask design. Fundamentally, we envision that this approach will be relevant not only for optics, but for any field where phase information needs to be extracted from intensity information.

### Funding:

This work was supported in part by the National Institute of General Medical Sciences Grant No. R35GM118067. PNP is a Xu Family Foundation Stanford Interdisciplinary Graduate Fellow.

### Acknowledgments:

We thank Anish R. Roy and Kayvon Pedram for stimulating discussions.